# Heat and mass transfer during crystal growing by the Czochralski method with a submerged vibrator

Fedyushkin A I

Ishlinsky Institute for Problems in Mechanics of the Russian Academy of Sciences

E-mail: fai@ipmnet.ru

Abstract. It was shown that the vibrations can intensify of heat- mass transfer and mixing of the melt, and also to reduce dynamic and temperature boundary layers and to increase a temperature gradient on melt/crystal interface, that can to increase rate of crystal growth by the Czochralski method with a submerged vibrator. It is can be very impotent for manufacturing and quality of crystals.

## 1. Introduction

The consequences of the vibration influences on the melt can have as a positive affect, improving the mixing of the melt, the heat removal from the crystal and the properties of the crystal, and can also have a negative affect, impairing the growth process and properties of the crystal. Therefore, for the vibration effect on the melt, it is necessary to choose certain amplitude-frequency parameters depending on the properties of the melt, the crystal growth method and the configuration of the vibration affect. Vibrational effects are diverse: they can be translational, rotational, vibrational, harmonic, nonlinear, etc. Many scientific works are devoted to the study of vibration flows, among which one can distinguish the work of Permian scientists [1], devoted to convective flows under vibration influences and their stability. In [1] the studies are based on the solution of equations for vibration deviations, which are obtained on the basis of averaging the Navier-Stokes equations by vibration parameters. In this paper, the simulation was carried out on the basis of solving the complete non-stationary Navier-Stokes equations, the results were processed over a long period of time and the average flow characteristics were determined.

The paper [2] provides examples of possible ways to control the melt mixing by gravity, rotation and vibration. The authors of [3] give an overview of the results of crystal growth under rotational-oscillatory and vibrational influence on the melt flow in order to control the crystal growth. The paper [4] presents the results of mathematical modeling of the effect of vibration effects on heat and mass transfer during crystal growth by Bridgman, Czochralski (with and without a submerged vibrator) and zone melting methods.

In the Czochralski method, the temperature distribution corresponds to an unstable hydrodynamic configuration, since the colder crystal is located above the melt. This suggests that the melt hydrodynamics and temperature distribution are more sensitive to hydrodynamic disturbances, such as vibrations. The influence of vibrational effects on the melt from the growing crystal was considered in [5]. However, the creation of controlled vibrations by a growing crystal is technologically more difficult than by a submerged vibrator. In this paper the influence of vibrations from the side of a vibrator immersed in the melt is considered. Modeling the processes of hydrodynamics and heat and mass transfer for the processes of crystal growth is a complex mathematical problem with determining the shape and dynamics of the crystallization interface. An overview of the results of mathematical modeling of heat and mass transfer in crystal growth and perspectives for the development of these works are given in [6]. This task, taking into account the vibration effects, can be even more complex, since it requires modeling of long-term unsteady processes with the resolution of high-frequency vibrations and possible changes in the computational domain.

In [7] for the Czochralski method with a submerged axial vibrator, the results of modeling the melt mixing process are presented and compared with experimental data.

In this paper we consider a simplified two-dimensional plane model of crystal growth by the Czochralski method with a submerged vibrator. The effect of vibrations on the width of the boundary layers near the crystallization front will be considered for different Prandtl numbers and vibrator locations. This effect of vibration influence on the width of the boundary layers was found by the author earlier for other crystal growth methods [4-5].

## 2. Problem statement and mathematical model

This article presents the results of the study of hydrodynamics and heat transfer for Czochralski crystal growth method with a submersible vibrator. The scheme of the computational domain is shown in Fig. 1. The computational domain is a square with sides L=H=3 cm Crystal with a diameter of d=1cm and immersed into the melt to a depth of 1mm, the vibrator has a diameter of 0.8 cm and thickness 1mm. It is assumed that the immersed vibrator is located under the crystal at a distance h horizontally and in the middle (parallel to the surface of the crystal). The removal of the surface of the vibrator from the crystal surface at a distance h=5mm, 8mm and 13.5 mm was considered. Irregular grids with refinement near the solid walls and the corners of the vibrator and the crystal were used in the calculations. The vibrator makes translational oscillatory movements along the vertical axis of the crystal according to the law: $y = y_0 + A\sin(2\pi ft)$, with frequency $f$ and small amplitude $A$ (it was equal $A = 100 \mu m$ or $A = 100 \mu m$), $y_0$ is initial location of vibrator. In this model the movement and rotations of the crystal and crucible are not considered.

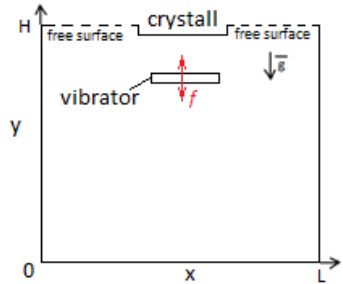

Fig. 1. Scheme of the computational domain.

Mathematical modeling of convective heat transfer is based on the numerical solution of the system of unsteady Navier-Stokes equations for incompressible fluid in the Boussinesq approximation (1). The simulation was carried out by two numerical methods: finite elements and control volumes, which are described in [4-5].

$$\frac{\partial V}{\partial t} + (V\nabla)V = -\frac{\nabla P}{\rho} + \nu\Delta V + \beta T\mathbf{g}, \quad divV = 0, \quad \frac{\partial T}{\partial t} + (V\nabla)T = a\Delta T \qquad (1)$$

where V is the velocity vector; P is the pressure; ρ is the density; T is the temperature; g is the free fall acceleration vector, $a$, $\nu$, $\beta$ are the coefficients of temperature conductivity, kinematic viscosity and linear temperature expansion, d is the crystal diameter, h is the distance from the crystal to the vibrator. An adhesion condition for the velocity at all solid boundaries is set subjected. At the free boundary is set slip condition or in the presence of thermo-capillary convection is set Marangoni force condition. The following conditions are set for the temperature: the crucible has a constant temperature in T=400K, the crystal (located at the top) has a temperature in T=300K, the thermal insulation condition is set on the free surface. At the initial moment, the liquid is assumed to be stationary with a constant temperature or conductivity regime distribution of temperature.

This problem is characterized by the following dimensionless numbers of Prandtl $Pr = v/a$, (or Péclet $Pe = Re\,Pr$), Grashof $Gr = g\beta\Delta TH^3/v^2$ (or Rayleigh $Ra = Gr\cdot Pr$), Marangoni $Ma = -d\sigma/dT \cdot \Delta TH /(v\rho a)$, vibrational Reynolds number $Re_{vibr} = 2\pi fAH/v$, the ratio $h/d$ of the distance from the crystal to the vibrator to the diameter of the crystal and the geometry of the vibrator.

In time, the numerical calculation was carried out in the absence of vibrations until reaching the stationary regime of fluid flow or up to several hundred periods of oscillation $\Theta = 1/f$ of the quasi-stationary regime in the case of vibration action. Then the results were averaged and analyzed. The averaged vibrational flow (AVF) $\Phi_{AVERAGE}$ was calculated in the process of averaging over time the numerical solutions $\Phi$ at the time interval (0,t) by the formula: $\Phi_{AVERAGE} = \frac{1}{t}\int_0^t \Phi d\tau$. The time long of averaging of flow and the temperature field depends on Prandtl number and, for example, for Pr=7 it must be at least $t > 400\Theta$, where $\Theta = 1/f$ - the period of vibration.

One of the difficulties of the numerical simulation of the steady AVF with a high frequency is the need to carry out calculations with a small time step $\Delta\tau$ (no more than $\Delta\tau \leq 0.01\Theta$), which requires a large time-consuming computer. A dynamic grid was used to simulate the motion of the submerged vibrator. Implicit conservative schemes were used in the calculations. However, the Courant number $C = 2\pi fA\Delta\tau/\Delta h$ and the Reynolds grid number $Re_c = 2\pi fA\Delta h/v$ were about 1.

## 3. Results

### 3.1. Flow and heat transfer without vibration

#### 3.1.1. Heat conductivity

The isotherms for the thermal conductivity mode (without vibrations and without convection) for the two cases (without and with a submerged vibrator, for Pr=7, h/d=0.5) are shown in Fig. 2a and in Fig. 2b respectively. Comparison of the temperature distribution for these cases shows that the existence of a thermally insulated stationary plate (no moving vibrator) under the crystal in the melt increases the width of the temperature boundary layer on crystal.

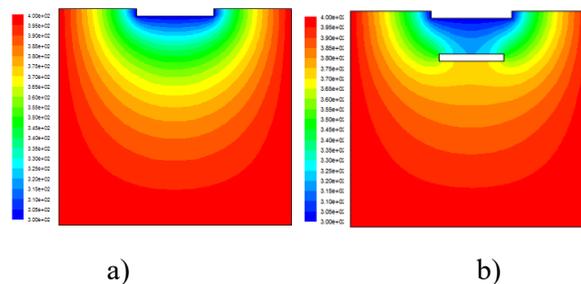

a) b)

Fig. 2. Isotherms at heat conductivity mode:
a) without vibrator, b) with submerged vibrator h/d=0.5.

#### 3.1.2. Natural and thermo-capillary convection

The effect of natural and thermo-capillary convection without a submersible vibrator on heat transfer for Rayleigh numbers from 0 to $10^6$ and Marangoni numbers from 0 to $10^6$ for Pr=7 (this value of the Prandtl number corresponds to the molten sodium nitrate) is considered.

In the case of natural convection only, the fluid flow has a lifting-lowering laminar character with a maximum velocity on the vertical axis. The calculation results showed that the maximum velocities

$\lambda \cdot (\partial T / \partial y)$) along the axis of the crucible for Rayleigh numbers (Ra=0 - $10^6$) confirm that with increasing Rayleigh number increases the module heat flux at the crystallization front. The thicknesses of the dynamic and temperature boundary layers are reduced proportionally to $Re^{-1/2}$ and $Pe^{-1/2}$, respectively.

In zero-gravity conditions (g=0), there may also be an intense flow in the melt caused by thermo-capillary convection generated on the free surface due to the dependence of the surface tension on the temperature. The intensity of thermo-capillary convection is determined by the value of the Marangoni number. Marangoni convection also creates a lift-down flow with two oppositely directed vortices, as in the case of only natural convection, but more intense than at the same values of the Rayleigh number. These flows are laminar and the maximum velocities are proportional to the number of Marangoni to the power of 1/2. The results of the calculations showed that the values of the Reynolds number calculated from the maximum flow rate for the Marangoni numbers Ma=10, $10^2$, $10^3$, $10^4$, $10^5$, $10^6$, respectively are equal: Re=0.2, 2.5, 31, 189, 684, 2612. In the case of thermo-capillary convection only the maximum of the velocity modulus is observed near the free surface, while in the case of natural convection the maximum of the velocity modulus is under the crystal in the center of the crucible. The thicknesses of the dynamic and thermal boundary layers under the crystal are refined with the increase of the Marangoni number, that could have an impact on the growth rate of the crystal.

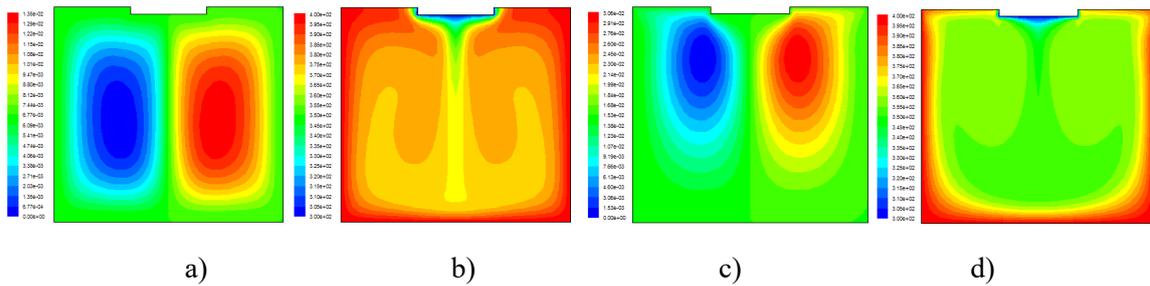

a)      b)      c)      d)

Fig. 3. Stream function (a, c) and isotherms (b, d)
for $Ra = 10^6, Ma = 0$ (a, b) and $Ra = 0, Ma = 10^5$ (c, d).

In Fig. 3 the results of modeling of natural and thermo-capillary convection in the form of stream function and isotherms for a liquid with Prandtl number $Pr = 7$ and for following Rayleigh and Marangoni numbers: $Ra = 10^6, Ma = 0$ (a, b) and $Ra = 0, Ma = 10^5$ (c, d) are shown. The structures of flows and temperature distributions at natural and thermo-capillary convection differ - this can explain by different mechanisms of flow generation. However, both convective flows have the same effect on the boundary layers under the crystal, although with different intensity.

*3.2. Influence of translational harmonic vibrations on heat transfer*
All the results presented below will apply to time-averaged temperatures and velocities for quasi-stationary fluid flow regimes

*The middle Prandtl number.*
Vibrations with small amplitude and frequency of several hertz can create a melt mixing, the intensity of which can be greater than the intensity of the convections. Therefore, it can be expected that the vibration effect can have a stronger effect on the boundary layers at the crystallization front.

If we consider the model described above with the following parameters: Pr=7, Ra=Ma=0, h/d=0.5, with vibrations of the vibrator with a frequency f=20Hz and amplitude A=0.4 mm, then this corresponds to the value of the vibrational Reynolds number equal 1500 ($Re_{vibr} = 1500$). In Fig. 4 for these parameters the results of calculations with and without vibration in the form of isotherms, tracks of flow and temperature profiles are presented.

The structure of the averaged flow is presented in Fig. 4c. It is shows how the vibrating immersed activator leads to the mixing of the entire volume of the melt. In Fig. 4d presents temperature profiles on vertical cross section (on axis) that show the effect of vibration on the temperature boundary layer and the temperature gradient near the crystallization front (Pr=7; $Re_{vibr}=1500$; h/d=0.5).

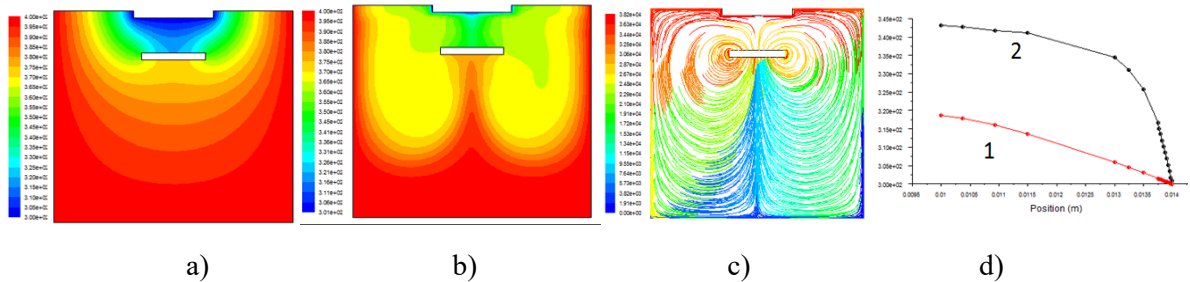

a)         b)         c)         d)

Fig. 4. Isotherms temperature: a) – without vibration, b) with vibrations c) flow tracks, d) temperature profiles on the axis section: curve 1 is without vibration, curve 2 is with vibration.

The effect of the distance of the vibrator from the crystal surface (for Pr=7, $Re_{vibr}=1500$, Ra=0, h/d=0.8 and h/d=0.5) is presented in Fig. 5. In Fig. 5a and Fig 5b are show isotherms for cases without vibrations and with vibrations for h/d=0.8. In Fig. 5c temperature profiles are presented on the vertical axis (middle vertical cross section) between the crystal and the vibrator for h/d=0.8 and h/d=0.5. The simulation results showed that the removal of the vibrator from the crystal at a distance h/d=0.8 reduces the effect on the vertical temperature gradient near the crystal, compared with the case of a closer arrangement of the vibrator h/d=0.5 (Fig. 5c).

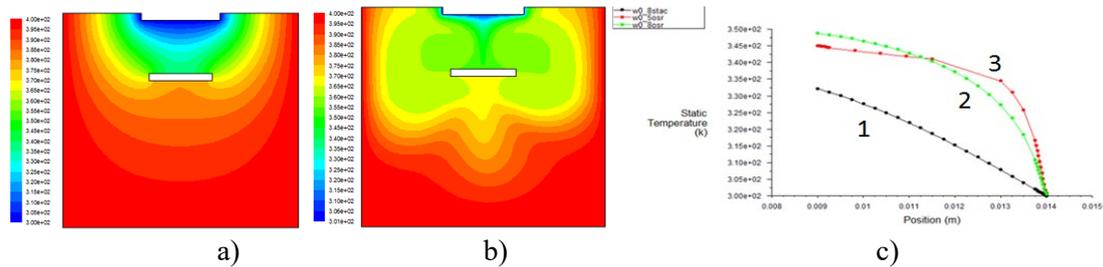

a)         b)         c)

Fig. 5. Isotherms (Pr=7, Ra=0, h/d=0.8), a) - without vibrations, b) - with vibrations $Re_{vibr}=1500$, c) temperature profiles (on middle vertical cross section): curve 1 is without vibrations (h/d=0.8), curve 2 is with vibration ($Re_{vibr}=1500$, h/d=0.8), curve 3 is with vibrations ($Re_{vibr}=1500$, h/d=0.5).

Consider the case of joint harmonic translational vibrations of a crystal and a submerged vibrator taking into account convection (Pr=7, $Re_{vibr}=1500$, $Ra=10^4$).

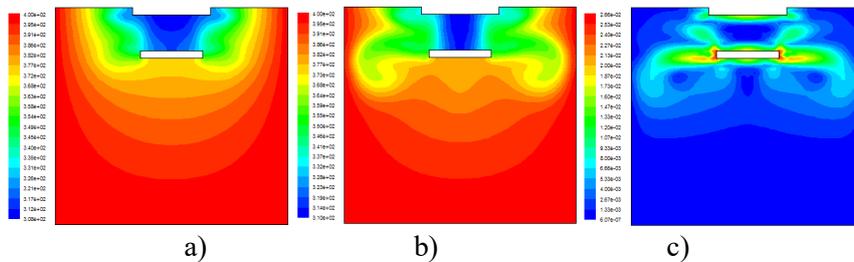

a)         b)         c)

Fig. 6. Isotherms at joint vibrations of the crystal and the submerged vibrator (Pr=7, $Re_{vibr}=1500$), a) - is without convection, b), c) - are with convection $Ra=10^4$; c) - the isolines of the module of velocity.

The combined action of the vibration of the crystal and the immersed vibrator with the same law synchronously and intensity corresponding to the Reynolds vibration number without convection is presented in Fig. 6a (without convection and vibration was presented in Fig. 1).

The combined action of the vibration of the crystal and the immersed vibrator in the presence of natural convection $Re_{vibr} = 1500$, $Ra = 10^4$, Pr=7 is presented in Fig. 6b,c. From the results presented in Fig. 6 it can be concluded that the joint synchronous vibrations of the crystal and the immersed vibrator made more homogenous of the temperature field under the crystal, and the natural convection contributes to an additional reduction in the thickness of the temperature boundary layer formed only by vibrations.

*The small Prandtl number.*

In this paragraph the simulation results for the molten metals (semiconductors) with the following dimensionless parameters: Pr=0.1, $Re_{vibr} = 1500$, Ra=0, h/d=0.5 will considered. This case is interesting because the rate of heat transfer in the melts of metals and semiconductors is very large and can be greater than or equal to the vibration velocities ($Af$). In Fig. 7, presented: a) - isotherms without vibration, b) - isotherms under vibration and c) - temperature profiles along the central vertical axis between the crystal and the vibrator (curve 1 - is without vibrations, curve 2 - is with vibrations). Due to the small Prandtl number, the effect of vibrations on the temperature boundary layer is insignificant (Fig. 7c).

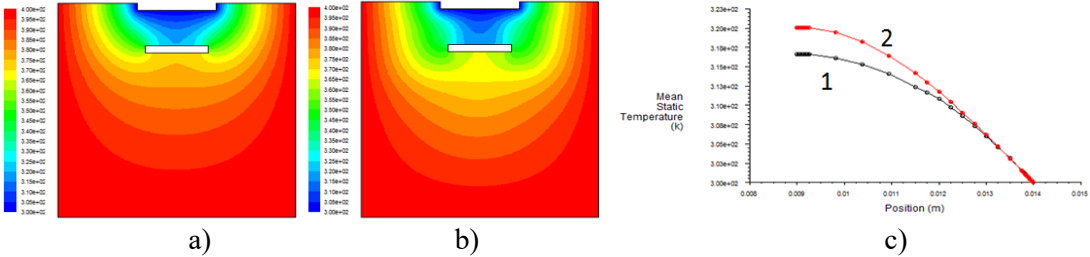

a)  b)  c)

Fig. 7. Isotherms (Pr=0.1, Ra=0, h/d=0.5, $Re_{vibr} = 1500$), a) – without vibrations, b) - with vibrations, c) temperature profiles in the central vertical section: 1 - is without vibrations, 2 - is with vibrations.

A stronger influence of the vibrations on the temperature boundary layer of molten metals (semiconductors) with small Prandtl numbers can be affected by increasing the amplitude and frequency of vibrations (vibrational Reynolds number $Re_{vibr}$) [4]. The selection of amplitude and frequency must be made from the condition that $A^2 f > a$.

*The large Prandtl number.*

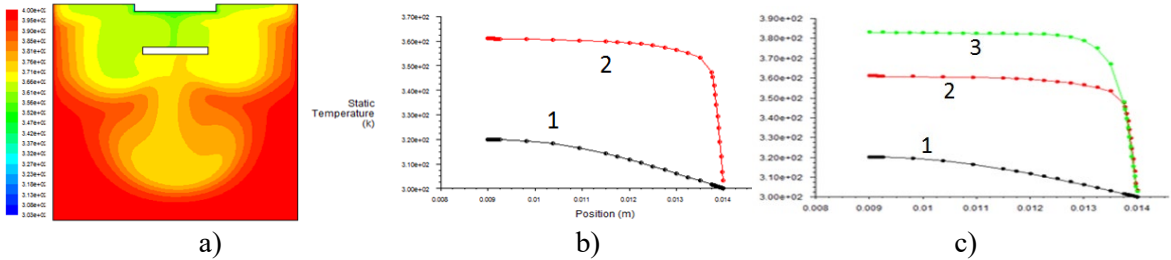

a)  b)  c)

Fig. 8. a) - Isotherms (Pr=100, Ra=0, h/d=0.5, $Re_{vibr} = 1500$), b) - profiles of the instantaneous temperature field in the middle vertical cross-section: curve 1 - without vibrations, curve 2 - with the vibrations, c) profiles of the averaged temperature in the middle vertical cross-section: curve 1 - without vibrations, curve 2 - with the vibrations when h/d=0.5, curve 3 - with the vibrations when h/d=1.35.

In this paragraph the simulation results for the melt of oxides with the following dimensionless parameters: Pr=100, $Re_{vibr}=1500$, Ra=0, h/d=0.5 are considered and presented in Fig. 8a,b.

The results on the effect of the distance of the vibrator from the crystal surface at Pr=100, Ra=0, h/d=1.35 and h/d=0.5 are presented in Fig. 8c. It can be concluded that for liquids with Pr=100 the changing of the distance of the vibrator from the crystal has less effect on the thermal field near the crystal than the changing of the distance of the vibrator for liquids with Pr=7 (Fig. 6b) (however the common effect of the vibrations on temperature layers stronger for Pr=100).

The effect of natural convection for our model to consider for fluid with Pr=100. In Fig. 9 presents the isotherms (Fig. 9a) and vertical profiles (Fig. 9b,c) of the averaged temperature fields obtained for Pr=100, $Re_{vibr}=1500$, $Ra=7\cdot10^5$, h/d=0.5, for time $t=600\Theta$. Numerical calculations are carried out with taking into account the presence of natural convection and without it, with the vibrations and without the vibrations. In Fig. 9b,c temperature profiles are presented for all these cases. The results show that natural convection, as well as vibration, contributes to an increase of the temperature gradient near the crystallization front (Fig. 9b, curve 2, Fig. 9c, curve 2). The vibration effect ($Re_{vibr}=1500$) has a stronger effect on reducing the thickness of the temperature boundary layer than natural convection ($Ra=7\cdot10^5$) (Fig. 9b,c).

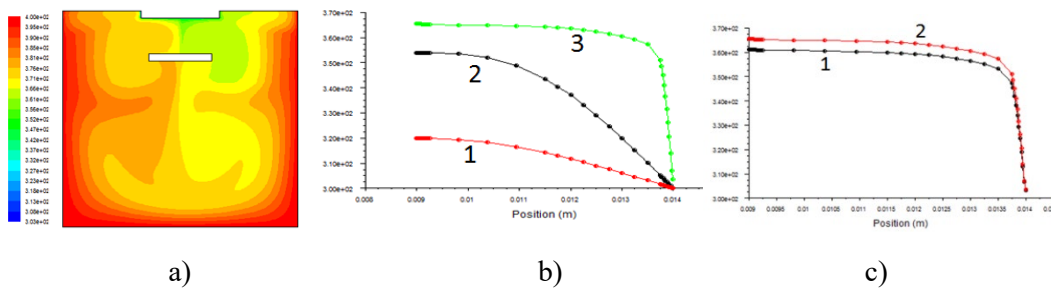

a)  b)  c)

Fig. 9. Isotherms (a) and temperature profiles (b, c) at Pr=100, Ra=7 $10^5$, h/d=0.5; b) - temperature profiles in the middle vertical cross-section: curve 1 - without vibrations and without convection, curve 2 – without vibrations, only with convection (Ra=7 $10^5$), curve 3 - with vibrations and convection ($Ra=7\cdot10^5$, $Re_{vibr}=1500$), c) - are vertical temperature profiles with vibrations ($Re_{vibr}=1500$): curve 1 - without convection (Ra=0), curve 2 - with convection ($Ra=7\cdot10^5$).

From the results presented in Fig. 4-9, it can be concluded that under the vibration action, the temperature boundary layer decreases, and this effect is the stronger the greater the value of Péclet number. To achieve the same effect in liquids with small Prandtl numbers (as is the case with large Prandtl numbers), it is necessary to increase the value of the vibrational Reynolds number.

*3.3. Rotationally oscillatory vibrations of the submerged plate*

Consider the rotational-oscillatory vibrations of the plate immersed in the melt, deviating at a small angle around the axis perpendicular to the plane of the surface of the computational domain and passing through the center of the plate. The scheme of the computational domain is shown in Fig. 10. Plate long 2 cm, thickness of 1 mm located at a distance of 1.45 cm from the surface of the crystal at the center of the computational domain. Rotational-oscillatory vibrations are performed at an angle $\varphi$ according to the harmonic law: $\varphi = A\cdot\sin(2\pi\omega\cdot t)$ with amplitudes A and frequency $\omega$=20 Hz, where $\varphi$ - the angle of deviation of the vibrator from the initial (horizontal) position shown in Fig. 10. Dimensionless vibrational Reynolds number is defined as $Re_{vibr}=2\pi\omega AH^2/\nu$.

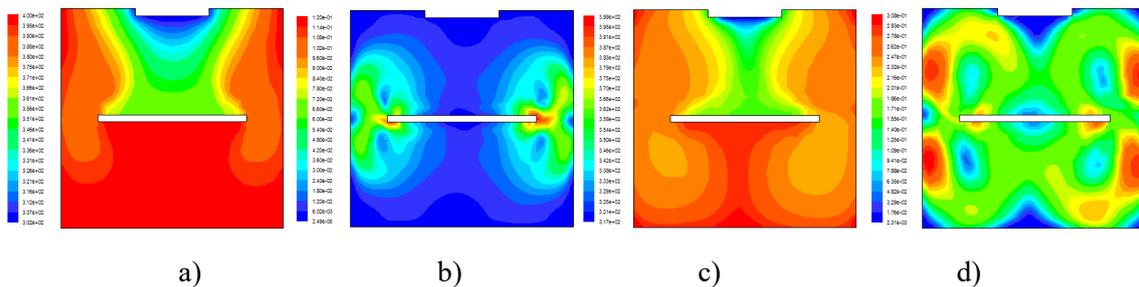

a)          b)          c)          d)

Fig. 10. Isotherms (a,c) and isolines of the velocity module (b, d) at rotational oscillation vibrations (Pr=7, $\omega = 20 Hz$) with amplitudes: $\omega = 20 Hz$, $A = 10^0$ (a, b) and $A = 25^0$ (c, d).

The results of numerical simulations showed that the rotational-oscillatory effects lead to the formation of four vortices in the structure of the averaged flow (Fig. 10). By controlling these vortices, it is also possible to change the structure of the flow and the temperature field near the crystallization front. The changing in the amplitude of the rotational-oscillatory vibrations $A$ from 10 to 25 degrees significantly changes the flow structure of the thickness of the boundary layers at the front of crystallization (Fig. 10).

**4. Conclusions**

For the model of the Czochralski crystal growth method with a submerged vibrator, for different properties of liquids with and without convection flows a decrease of the thickness of the boundary layers under vibration are shown, which confirms the general character of this fact [4-5]. This effect is stronger with increasing Prandtl number. To influence of vibration on the temperature layer we have to select of amplitude and frequency from the condition that $A^2 f > a$.

Rotationally oscillatory vibrations of the submerged plate well intensify the flow and mixing of the melt, affect on the averaged temperature and the structure of the boundary layers under the crystal, so it can also be used as a control mechanism.